\documentclass[twocolumn,showpacs,preprintnumbers]{revtex4}%
\usepackage{amsmath}
\usepackage{graphicx}
\usepackage{dcolumn}
\usepackage{bm}%
\usepackage{amsfonts}%
\usepackage{amssymb}
\begin{document}
\title{Bistability in Feshbach Resonance}
\author{Hong Y. Ling}
\affiliation{Department of Physics and Astronomy, Rowan University, Glassboro, New Jersey,
08028-1700, USA}

\begin{abstract}
A coupled atom-molecule condensate with an intraspecies Feshbach resonance is
employed to explore matter wave bistability both in the presence and in the
absence of a unidirectional optical ring cavity. In particular, a set of
conditions are derived that allow the threshold for bistability, due both to
two-body s-wave scatterings and to cavity-mediated two-body interactions, to
be determined analytically. \ The latter bistability is found to support, not
only transitions between a mixed (atom-molecule) state and a pure molecular
state as in the former bistability, but also transitions between two distinct
mixed states.

\end{abstract}
\date{\today }
\maketitle

\section{Introduction}

The subject of optical bistability \cite{gibbs85}, to which Lorenzo Narducci
contributed greatly during his prime years of life, was brought to spotlight
again by recent experimental demonstration of optical bistability in a
microcavity with a cavity field at the level of a single photon
\cite{kurn07,esslinger08}. \ Instead of thermal gases, as typically employed
by Lorenzo's generation, where the thermal de Broglie's wavelength of the
particles is far smaller than the interparticle spacing, more recent trend in
cavity quantum electrodynamics (QED) focuses on cavity systems with ultracold
quantum gases \cite{kurn07,esslinger08}, opening the door to studies such as
cavity-mediated bistable Mott-insulator to superfluid phase transition
\cite{lewenstein08,chen09}, where the particle statistical nature becomes an
essential feature of the cavity problem. \ In contrast to 1980s, when the
surge of interest in optical bistability was inspired by the prospect of its
use as a switch in an all-optical computer \cite{gibbs85}, the current surge
of interest in the same topic was, however, motivated largely by the
equivalence of the cavity condensate system \cite{esslinger08} to the cavity
opto-mechanical system \cite{braginsky80}. \ The study of the latter falls
into the realm of cavity optomechanics, a rapidly emerging field which aims to
use the cavity-assisted radiation force \cite{ritsch00} to cool mechanical
device, ranging from nano- or micro-mechanical cantilevers to macroscopic
mirrors in LIGO project, down ultimately to their quantum mechanical ground
state \cite{girvin09}.%
\begin{figure}
[ptb]
\begin{center}
\includegraphics[
height=1.5708in,
width=3.4587in
]%
{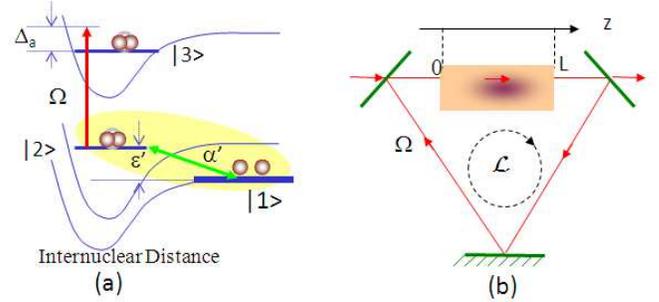}%
\caption{(Color online) (a) The schematic energy diagram (with the shaded part
representing the intraspecies Feshbach process) for (b) a unidirectional ring
cavity + coupled homonuclear atom-molecule condensate system.}%
\label{Fig:schematic}%
\end{center}
\end{figure}

In this paper, we focus on a cavity system containing a coupled homonuclear
atom-molecule condensate as illustrated in Fig. \ref{Fig:schematic}. \ In the
absence of the cavity, molecular state $\left\vert 2\right\rangle $ is coupled
only to free atomic state $\left\vert 1\right\rangle $ by an intraspecies
Feshbach resonance \cite{feshbach} characterized with a strength
$\alpha^{\prime}$ and a detuning $\epsilon^{\prime}$ [the shaded part in Fig.
\ref{Fig:schematic}(a)]. \ In the cavity QED setting, state $\left\vert
2\right\rangle $ is coupled, besides to state $\left\vert 1\right\rangle $,
also to excited molecular state $\left\vert 3\right\rangle $ by a
unidirectional ring cavity of a total length $\mathcal{L}$. \ The cavity is
driven by an external laser of wavenumber $k$, polarization $\mathbf{\hat
{\varepsilon}}$, and frequency $\omega$, which is tuned far away from the
$\left\vert 3\right\rangle \leftrightarrow\left\vert 2\right\rangle $
molecular transition. \ In addition, the cavity is assumed to possess a
sufficiently large intermode spacing $\mathcal{L}/c$ (with $c$ being the light
speed in vacuum) so that only the cavity mode with a longitudinal mode
frequency $\omega_{c}$ closest to $\omega$ is relevant to our study, where
$\omega_{c}\mathcal{L}/c$ equals integer multiple of $2\pi$. \ Further, the
cavity mode is assumed to overlap with the condensate in a spatial region,
characterized with a length $L$ and a cross-sectional area $A$, large enough
so that the condensate can be treated as a uniform system with an effective
volume $V_{a}=LA$ and a total atom number $N_{a}$ (counting those in
molecules). \ 

A similar cavity + condensate system has recently been studied \cite{search07}%
, in which atoms are converted into molecules by photoassociation as opposed
to magnetoassociation (Feshbach resonance) in our model. \ There, absorption
of a cavity photon will convert two atoms into a molecule, and emission of a
cavity photon will dissociate two atoms into a\ molecule. \ This is
reminiscent of the absorptive bistability in a driven optical cavity
containing an ensemble of two-level atoms. In contrast, the cavity field in
our model is to introduce a phase shift to the molecular field, not the
population exchange between atoms and molecules. \ As a result, the
bistability in our model is of dispersive nature. \ Further, interaction
between electronic dipoles and a cavity field in Ref. \cite{search07} gives
rise to atom-molecule coupling, which, since the cavity field is itself a
dynamical variable, changes with time. In contrast, in our model, it is the
hyperfine interaction - the coupling between electron spins and nuclear spins
of two colliding atoms that result in atom-molecule coupling, which is
therefore fixed by the atomic internal structure, independent of cavity field.
\ In this respect, our model is analogous more closely to a cavity system with
a spin-1 condensate (also recently proposed \cite{zhoulu09}) than to a cavity
system with photoassociation \cite{search07}; atom exchanges among different
internal states in the spin-1 condensate are accomplished by spin-exchange
interaction, which is also independent of cavity field. \ 

The interest in Feshbach resonance stems primarily from its use as an
effective tool to coherently create molecular BECs from the existing atomic
BECs. \ In a typical experiment, \ conversion of atomic BECs into molecular
BECs is carried out by ramping the Feshbach detuning across the resonance.
\ This method relies on the existence of a mixed (or dressed) atom-molecule
state \cite{stoof04a}, and the ability of this state to change its composition
from predominantly atomic to predominantly molecular species when the Feshbach
detuning $\epsilon^{\prime}$ is tuned from above to bellow the resonance.
\ The question that we want to pursue, in this paper, is how the molecular
population in state $\left\vert 2\right\rangle $ can be made to vary with the
Feshbach detuning in a bistable fashion, instead of monotonously as in typical
situations. \ A bistable crossover adds new meaning to the Feshbach resonance:
whether the system is in the atomic or molecular extreme is determined not
only by the Feshbach detuning but also by the history of the system. \ 

To this end, we first formulate, in Sec. II, a semiclassical mean-field
description of our model in the limit where both the excited molecular field
and the optical cavity field can be adiabatically eliminated. \ We then
explore the matter-wave bistability due to the two-body s-wave collisions in
Sec. III A and that due to the cavity-mediated effective interaction between
two Feshbach molecules in Sec. III B. \ Finally, we provide a summary in Sec. IV.

\section{The Basic Equations}

In this section, we take a semiclassical approach, in which optical fields are
treated classically while matter fields are treated quantum mechanically, to
formulate a theoretical description of the proposed cavity + condensate
system. \ This is the same approach that Lorenzo embraced in many of his works
\cite{lasers}, except that quantization is now performed at a level for a
many-body system, instead of a single-body system as in\ a typical
semiclassical laser theory. \ To begin with, we expand optical field
$\mathbf{E}\left(  \mathbf{r},t\right)  $ in terms of the slowly varying
amplitude $F$ in space $\mathbf{r}$ and time $t$ according to
\begin{equation}
\mathbf{E}\left(  \mathbf{r},t\right)  =\frac{1}{2}F\mathbf{\hat{\varepsilon}%
}e^{i\left(  kz-\omega t\right)  }+c.c,\label{E}%
\end{equation}
and matter field $\hat{\psi}\left(  \mathbf{r},t\right)  $ in terms of
$\hat{\psi}_{i}=\hat{c}_{i}/\sqrt{N_{a}}$ according to%
\begin{equation}
\hat{\psi}\left(  \mathbf{r}\text{,}t\right)  =\sqrt{n_{a}}\left[  \hat{\psi
}_{1}\left\vert 1\right\rangle +\hat{\psi}_{2}\left\vert 2\right\rangle
+\hat{\psi}_{3}e^{i\left(  kz-\omega t\right)  }\left\vert 3\right\rangle
\right]  ,\label{psiField}%
\end{equation}
where $\hat{c}_{i}$ denotes the operator in momentum space for annihilating a
bosonic particle in state $\left\vert i\right\rangle $ and $n_{a}=N_{a}/V_{a}$
is the total atom number density. \ The expansion in Eq. (\ref{psiField}) is
carried out in a frame rotating at the laser frequency $\omega$. \ In arriving
at Eq. (\ref{psiField}), we have assumed that the particles in states
$\left\vert 1\right\rangle $ and $\left\vert 2\right\rangle $ are all
condensed to their respective zero momentum modes, and those in state
$\left\vert 3\right\rangle $ to the mode with $\hbar k$ momentum in accordance
with momentum conservation during photon-atom interaction.

The coupled atom-molecule system, within these approximations, is then
described by the following Hamiltonian
\begin{align}
\hat{H}/\hbar N_{a}  & =\epsilon^{\prime}\hat{\psi}_{2}^{\dag}\hat{\psi}%
_{2}+\sqrt{n_{a}}\left(  \frac{\alpha^{\prime}}{2}\hat{\psi}_{2}^{\dag}%
\hat{\psi}_{1}^{2}+h.c\right) \nonumber\\
& +n_{a}\frac{\chi_{ij}^{\prime}}{2}\hat{\psi}_{i}^{\dag}\hat{\psi}_{j}^{\dag
}\hat{\psi}_{j}\hat{\psi}_{i}\text{ }\left(  i,j=1\text{ or }2\right)
\nonumber\\
& +\left(  \epsilon^{\prime}-\Delta_{a}\right)  \hat{\psi}_{3}^{\dag}\hat
{\psi}_{3}-\left(  \Omega\hat{\psi}_{3}^{\dag}\hat{\psi}_{2}+h.c\right)
,\label{Hamiltonian1}%
\end{align}
where repeated indices are to be summed from 1 to 2. \ In Eq.
(\ref{Hamiltonian1}), the first line describes the Feshbach resonance of
strength $\alpha^{\prime}$, the second line the s-wave collisions of strength
$\chi_{ij}^{\prime}$ $(=\chi_{ji}^{\prime})$ between states $\left\vert
i\right\rangle $ and $\left\vert j\right\rangle $, and the last line the part
of Hamiltonian involving excited state $\left\vert 3\right\rangle $. In the
last line, the first term denotes the energy of state $\left\vert
3\right\rangle $ in the rotating frame, where $\Delta_{a}$ is the laser
detuning, and the second term stands for the laser-induced electric dipole
interaction, where $\Omega=\mu_{32}F/2\hbar$ is the Rabi frequency, $\mu
_{32}=\left\langle 3\right\vert \mathbf{\hat{\mu}}\cdot\mathbf{\hat
{\varepsilon}}\left\vert 2\right\rangle $ the matrix element, and
$\mathbf{\hat{\mu}}$ the electric dipole moment operator. \ Finally,
collisions involving the final excited state $\left\vert 3\right\rangle $ are
ignored since state $\left\vert 3\right\rangle $ remains virtually empty in
our model. \ 

An important concept in the semiclassical approach is the macroscopic
polarization defined as $\mathbf{P}\left(  \mathbf{r},t\right)  =$
$\left\langle \hat{\psi}^{\dag}\left(  \mathbf{r}\text{,}t\right)
\mathbf{\hat{\mu}}\hat{\psi}\left(  \mathbf{r}\text{,}t\right)  \right\rangle
$. \ This polarization is found, when the use of both Eq. (\ref{psiField}) and
the selection rule are made, to possess the same mathematical form as the
optical field in Eq. (\ref{E}),%
\begin{equation}
\mathbf{P}\left(  \mathbf{r},t\right)  =\frac{1}{2}P\mathbf{\hat{\varepsilon}%
}e^{i\left(  kz-\omega t\right)  }+c.c.,\label{polarization}%
\end{equation}
where $P=2n_{a}\mu_{32}\left\langle \hat{\psi}_{2}\hat{\psi}_{3}^{\ast
}\right\rangle $ represents the slowly varying part of the polarization. The
evolution of the optical field is then governed by the Maxwell's equation,
which, under the slowly varying envelope approximation, takes the form
\begin{equation}
c\frac{\partial\Omega}{\partial z}+\frac{\partial\Omega}{\partial t}%
=i\frac{\mu_{0}\omega c}{4\hbar}\mu_{32}P\text{,}\label{Maxwell}%
\end{equation}
where $\mu_{0}$ $\left(  \epsilon_{0}\right)  $ is the magnetic (electric)
permeability in vacuum. \ Equation (\ref{Maxwell}) clearly shows that
polarization plays the role of a bridge between the classical optical field in
Eq. (\ref{E}) and the quantum matter fields in Eq. (\ref{psiField}). \ 

The final component in the semiclassical approach pertaining to any QED
problems is the boundary condition, which, in our case and under the
assumption that both input and output mirrors\ have the same reflectivity $R$
(transmissivity $T=1-R$), can be cast into the form \cite{orozco89}
\begin{equation}
\Omega\left(  0,t\right)  =TY+R\Omega\left(  L,t-\Delta t\right)
e^{i\Delta_{c}\mathcal{L}/c},\label{boundary}%
\end{equation}
where $Y$ is the scaled amplitude of the incident field, $\Delta t=\left(
\mathcal{L}-L\right)  /c$ the transit free propagation time inside the cavity,
and $\Delta_{c}=\omega-\omega_{c}$ the cavity mode frequency detuning.
\ Equation (\ref{boundary}) links the field entering $\left(  z=0\right)  $ to
that leaving the condensate $\left(  z=L\right)  $, and hence implements the
concept of feedback, which is the most important feature of any optical
cavities. \ The school led by Bonifacio, Lugiato, and Narducci distinguishes
itself by its insistence on a rigorous (first-principle) treatment of the
boundary condition - a treatment made up of both a transformation mapping two
non-isochronous events in Eq. (\ref{boundary}) into two isochronous events
\cite{lugiato85} and a set of conditions embodying the notion of
\textquotedblleft mean-field limit\textquotedblright\ \cite{lugiato84}.
\ Following this treatment, we explicitly eliminate the spatial derivative in
Eq. (\ref{Maxwell}), rendering Eq. (\ref{Maxwell}) into
\begin{equation}
\frac{d\Omega}{dt}=\left(  i\Delta_{c}-\kappa\right)  \Omega+\kappa
Y+ig^{2}N_{a}\left\langle \hat{\psi}_{2}\hat{\psi}_{3}^{\dag}\right\rangle
,\label{cavity field equation}%
\end{equation}
where $\kappa=c\left\vert \ln R\right\vert /\mathcal{L}$ is the cavity damping
rate, $g=\mu_{32}\sqrt{\omega/2\hbar\epsilon_{0}V_{c}}$ the \textquotedblleft
Rabi frequency\textquotedblright\ per photon, and $V_{c}=\mathcal{L}A$ the
total cavity volume.

In this paper, we restrict our study to the parameter regime in which both the
time scale for the optical field (on the order of $1/\kappa$) and that for the
molecular field of state $\left\vert 3\right\rangle $ (on the order of
$1/\Delta_{a}$) are far shorter than those for the Feshbach degrees of
freedom, namely, the matter fields corresponding to states $\left\vert
1\right\rangle $ and $\left\vert 2\right\rangle $. \ This restriction allows
us to adiabatically eliminate the former fast variables in favor of the latter
slow ones, reducing, from the Heisenberg's equations for $\hat{\psi}_{i}$ and
the Maxwell's equation for $\Omega$, a set of equations involving only the
slow degrees of freedom $\psi_{1}$ and $\psi_{2}$: \
\begin{subequations}
\label{two fields}%
\begin{align}
i\frac{d\psi_{1}}{dt}  & =\chi_{1i}n_{i}\psi_{1}+\alpha\psi_{2}\psi_{1}^{\ast
},\\
i\frac{d\psi_{2}}{dt}  & =\left[  \epsilon^{\prime}+\chi_{2i}n_{i}+F\left(
n_{2}\right)  \right]  \psi_{2}+\frac{\alpha}{2}\psi_{1}^{2}.
\end{align}
where we have adopted the standard mean-field theory, treating $\hat{\psi}%
_{i}$ as $c$-numbers $\psi_{i}$. \ In Eqs. (\ref{two fields}), $n_{1}%
=\left\vert \psi_{1}\right\vert ^{2}$, $n_{2}=\left\vert \psi_{2}\right\vert
^{2}$, $\alpha=\alpha^{\prime}\sqrt{n_{a}}$, and $\chi_{ij}=\chi_{ij}^{\prime
}n_{a}$ are the renormalized quantities, and
\end{subequations}
\begin{equation}
F\left(  n_{2}\right)  =\frac{\kappa^{2}\eta N_{c}}{\kappa^{2}+\left(
\Delta_{c}-\eta N_{a}n_{2}\right)  ^{2}},\label{f(n2)}%
\end{equation}
represents the phase shift to the molecular field induced by the cavity
optical field, where $N_{c}=Y^{2}/g^{2}$ represents the number of injected
intracavity photons and $\eta=g^{2}/\Delta_{a}$ measures the effective
atom-cavity coupling strength. (In this paper,without loss of generality, we
limit $\Delta_{a}>0$ so $\eta$ is always positive.)

\section{Discussions}

\ In our system, stationary solutions can be divided into mixed atom-molecule
states in which each species has a \textit{finite} density $n_{i}$, and pure
molecular or atomic phase in which one of species has a zero density. \ In
this section, we study bistability in the Feshbach process by focusing on the
mixed atom-molecule state where each species can be described by the field
$\psi_{i}=\sqrt{n_{i}}e^{i\theta_{i}}$ characterized with a well defined phase
$\theta_{i}$. (Note that such a description cannot be applied to pure
molecular or atomic phases as the phase associated with an empty component is
not defined.)\ \ \ For such a state, we can apply the total atom number
conservation $n_{1}+2n_{2}=1$ and simplify Eqs. (\ref{two fields}) into a set
of equations involving only two variables - a molecular density $n\equiv
n_{2}$ and a phase mismatch $\theta=\theta_{2}-2\theta_{1}$. \ This set of
equations, in a unit system in which $\epsilon=$ $(\epsilon^{\prime}-\chi
_{12}-2\chi_{11})/\alpha$ is defined as the effective Feshbach detuning,
$\chi=[\chi_{22}+4\left(  \chi_{11}-\chi_{12}\right)  ]/\alpha$ as the
effective Kerr nonlinear coefficient for the molecular field, $\delta
=\Delta_{c}/\kappa$ as the cavity detuning, and $\tau=\alpha t$ as the time,
take the form%

\begin{subequations}
\label{two variable equation}%
\begin{align}
\frac{dn}{d\tau}  & =-\left(  1-2n\right)  \sqrt{n}\sin\theta,\\
\frac{d\theta}{d\tau}  & =\epsilon+\chi n+\frac{1}{2}\frac{1-6n}{\sqrt{n}}%
\cos\theta+f\left(  n\right)  ,
\end{align}
where
\end{subequations}
\begin{equation}
f\left(  n\right)  =2\frac{B/C}{1+\left(  \delta-Cn\right)  ^{2}%
}.\label{f scaled}%
\end{equation}
In arriving at Eqs. (\ref{two variable equation}), we have replaced $N_{a}$
and $N_{c}$ in Eq. (\ref{f(n2)}) in favor of two unitless parameters,
\begin{equation}
C=\eta N_{a}/\kappa=g^{2}N_{a}/\Delta_{a}\kappa,\label{C}%
\end{equation}
and
\begin{equation}
B=\eta^{2}N_{c}N_{a}/\left(  2\kappa\alpha\right)  .\label{B}%
\end{equation}
In contrast to $C$, which is bose enhanced by atom number only, $B$ is bose
enhanced by both photon and atom numbers. \ Note that when $\Delta_{a}$ is
replaced with the decay rate of the excited state $\left\vert 3\right\rangle
$, $C$ \ becomes the so-called atomic \textquotedblleft cooperative
parameter\textquotedblright\ \cite{kimble94}. \ As we will see shortly,
cavity-mediated bistability depends crucially on the values of $C$ and $B$. \ 

As in other studies \cite{zhou07,radzihovsky08}, Eqs.
(\ref{two variable equation}) of the type including Feshbach resonance support
two branches of steady-states: one with $\theta=0$ and the other with
$\theta=\pi$. \ This feature is expected since at zero temperature the
intraspecies Feshbach resonance represents a matter wave analog of the second
harmonic generation in nonlinear optics where the phase-matching condition
plays an important role. \ In this paper, without the loss of generality, we
take $\alpha>0$. Under such a circumstance, the branch\ with $\theta=\pi$ not
only always has a lower energy than the branch with $\theta=0$, but also has
the property of consisting primarily atom species in the limit of a large
positive Feshbach detuning. \ For these reasons, we will focus on the branch
with $\theta=\pi$, determined at steady state by%
\begin{equation}
\epsilon=-\chi n+\frac{1}{2}\frac{1-6n}{\sqrt{n}}-f\left(  n\right)
,\label{epsilon steady state}%
\end{equation}
where for notational simplicity, same symbols are used to stand for the steady
state variables.

To carry out the stability analysis, we apply the standard linearization
procedure that Lorenzo used extensively in the context of his interest in
laser instabilities \cite{narducci88,abraham85}, and derive from Eqs.
(\ref{two variable equation}) a set of linearly coupled equations
\begin{equation}
\frac{d}{d\tau}\left(
\begin{array}
[c]{c}%
\delta n\\
\delta\theta
\end{array}
\right)  =\left[
\begin{array}
[c]{cc}%
0 & \left(  1-2n\right)  \sqrt{n}\\
-\frac{d\epsilon}{dn} & 0
\end{array}
\right]  \left(
\begin{array}
[c]{c}%
\delta n\\
\delta\theta
\end{array}
\right)  ,\label{linear stability}%
\end{equation}
where $\delta n$ and $\delta\theta$ are small departures from the
corresponding steady state variables, and $\epsilon$ is given by Eq.
(\ref{epsilon steady state}). \ The eigenvalues of Eqs.
(\ref{linear stability}) are then found to take two values: $\pm\sqrt{-\left(
1-2n\right)  \sqrt{n}d\epsilon/dn}$, which, since $n<0.5$, clearly indicate
that only when $d\epsilon/dn>0$ or equivalently when $dn/d\epsilon$ $>0$ can
the eigenvalues become complex. \ In another words, any state at which the
slope of $n$ as a function of $\epsilon$ is positive is unstable. \ \ Note
that such a conclusion may not hold, if we do not impose, in the previous
section, the conditions that allow $\psi_{3}$ and $\Omega$ to be adiabatically
eliminated. \ In the case of absorptive optical bistability, it is well known
that the upper branch which is stable according to the slope criterion may
become unstable; instability there is manifested in the form of self-pulsings
\cite{bonifacio78}.

Thus, we see that the stability analysis here amounts to analyzing the points
at which $d\epsilon/dn=0$, which, by definition, simply corresponds to the
critical transition points in a bistable (or multi-stable) system. \ \ As a
result, in what follows, we carry out bistability study by focusing on\ the
equation
\begin{equation}
d\left(  n\right)  =\chi-h\left(  n\right)  ,\label{depsilon}%
\end{equation}
derived from Eq. (\ref{epsilon steady state}) under the condition that
$d\epsilon/dn=0$, where%
\begin{align}
d\left(  n\right)   & =-4B\frac{\left(  \delta-Cn\right)  }{\left[  1+\left(
\delta-Cn\right)  ^{2}\right]  ^{2}},\label{df/dn2}\\
h\left(  n\right)   & =-\frac{3}{2}n^{-1/2}-\frac{1}{4}n^{-3/2}.\label{h}%
\end{align}

\subsection{Collision-Induced Bistability}

\begin{figure}
[ptb]
\begin{center}
\includegraphics[
height=1.5741in,
width=3.32in
]%
{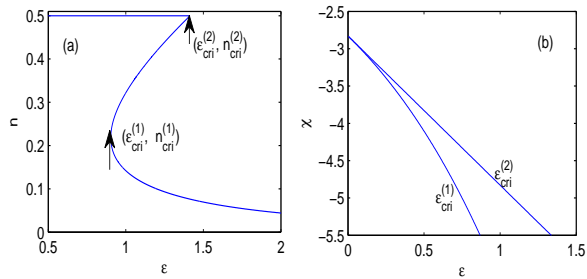}%
\caption{(a) Molecular population $n$ as a function of $\epsilon$ when
$\chi=2\chi_{th}=-4\sqrt{2}$. \ (b) $\epsilon_{cri}^{\left(  1\right)  }$ and
$\epsilon_{cri}^{\left(  2\right)  }$ at different values of $\chi$.}%
\label{Fig:bistability1}%
\end{center}
\end{figure}

In our model, the molecular field (at state $\left\vert 2\right\rangle $) is
subject to two types of self phase modulations: one, described by a Kerr type
of nonlinear term $\chi n,$ originates from short-range s-wave scatterings,
and another, described by $f\left(  n\right)  $ in Eq. (\ref{f scaled}), stems
from cavity-mediated long-range two-body collisions. \ In order to
differentiate their roles in the formation of matter wave bistability, we
first remove the cavity component and study the bistability due solely to the
Kerr nonlinearity by solving
\begin{equation}
\chi=h\left(  n\right)  ,\label{chi no cavity}%
\end{equation}
obtained from Eq. (\ref{depsilon}) by setting $d\left(  n\right)  =0$. \ As
one may easily verify, $h\left(  n\right)  $ in Eq. (\ref{h}) is a
monotonously increasing function of $n$. \ As a result, in order for
Eq.(\ref{chi no cavity}) to have real roots within $n<0.5$, $\chi$ must be
less than $h\left(  n=0.5\right)  $ or
\begin{equation}
\chi\leq\chi_{th}\equiv-2\sqrt{2}.\label{bistability condition f=0}%
\end{equation}
This condition is similar to that of Ref. \cite{jiang09} for a heteronuclear
atom-molecule system with an interspecies Feshbach resonance \cite{zhou07}. It
simply reflects the fact that for bistability to take place, there must be a
sufficiently strong positive feedback between the molecular population and the
effective Feshbach detuning $\epsilon+\chi n$. \ A negative $\chi$ fulfills
this positive feedback; as can be seen, with a negative $\chi$, an increase in
the molecular density decreases\ the effective detuning, which, in turn,
further increases the molecular density, or vice versa. \ Such a chain of
positive reaction under condition (\ref{bistability condition f=0}) can lead
to the formation of the critical transition points around which the molecular
population changes in a runaway fashion. \ Indeed, under condition
(\ref{bistability condition f=0}), Eq. (\ref{depsilon}) is found to support a
critical point with a critical molecular density $n_{cri}^{\left(  1\right)
}$ given by,
\begin{align}
\sqrt{n_{cri}^{\left(  1\right)  }}  & =\left(  \frac{1}{16\left\vert
\chi\right\vert }\right)  ^{\frac{1}{3}}\left(  \sqrt{1+\frac{2}{\chi^{2}}%
}+1\right)  ^{\frac{2}{3}}+\nonumber\\
& \left(  \frac{1}{16\left\vert \chi\right\vert }\right)  ^{\frac{1}{3}%
}\left(  \sqrt{1+\frac{2}{\chi^{2}}}-1\right)  ^{\frac{2}{3}}-\frac{1}{2\chi
},\label{n^(1)}%
\end{align}
and a critical Feshbach detuning $\epsilon_{cri}^{\left(  1\right)  }$
determined by Eq. (\ref{epsilon steady state}) when $n$ is replaced with
$n_{cri}^{\left(  1\right)  }$ in Eq. (\ref{n^(1)}). \ Figure 2(a) shows a
typical example of bistability (with $\chi=2\chi_{th}$). \ In addition to the
mixed state, there is a pure molecular state with $n=0.5$ obtained from Eqs.
(\ref{two fields}), not from Eqs. (\ref{two variable equation}) which, by
definition, only holds for mixed states. \ (Note that the pure atomic state is
prohibited by the nature of intraspecies Feshbach resonance
\cite{radzihovsky04,stoof04,radzihovsky08}.) \ The interception between this
pure state and the mixed atom-molecular state defines the second critical
point
\begin{equation}
n_{cri}^{\left(  2\right)  }=0.5,\epsilon_{cri}^{\left(  2\right)  }%
=-\frac{\chi}{2}-\sqrt{2}.
\end{equation}
The threshold for bistability is reached when the two critical points become
degenerate. \ Clearly, this happens at $n=0.5$ when $\chi=\chi_{th}$. \ Figure
\ref{Fig:bistability1}(b) shows that when $\chi$ is bellow $\chi_{th}$, the
size of the hysteresis loop (measured by $\epsilon_{cri}^{\left(  2\right)
}-$ $\epsilon_{cri}^{\left(  1\right)  }$) increases with $\left\vert
\chi\right\vert $. \

To see the implication of condition (\ref{bistability condition f=0}) to a
realistic system, consider, for example, the Feshbach resonance located at the
magnetic field $85.3$ mT in $^{23}$Na \cite{julienne00,ling02}. \ This
resonance has a width of $0.95\mu$T or equivalently a Feshbach coupling
strength of $\alpha^{\prime}=4.22\times10^{-6}m^{3/2}s^{-1}$. \ \ Taking the
total atom number density to be $10^{20}$ m$^{-3}$ and using 3.4 nm as the
s-wave scattering length for sodium atoms \cite{verhaar99}, we find that
$\chi_{11}=$ 1.18$\times10^{4}$ s$^{-1}$ and $\alpha=$4.22 $\times10^{4}$
s$^{-1}=3.58\chi_{11}$. If we further assume $\chi_{22}=\chi_{11}$, we see
from Eq. (\ref{bistability condition f=0}) that this requires $\chi_{12}%
>\frac{1}{4}\left(  \chi_{22}+4\chi_{11}+2\sqrt{2}\alpha\right)
>3.77\chi_{11}$. \ 

\subsection{Cavity-Mediated Bistability}

The above example means to illustrate that Eq.
(\ref{bistability condition f=0}) can be fulfilled only when both the total
atom density and the two-body interspecies collisional strength are
sufficiently large, a condition which is difficult to meet under typical
systems. \ In this subsection, we turn our attention to the cavity model in
Fig. \ref{Fig:schematic}, and pursue, from Eq. (\ref{depsilon}), the question
of under what cavity parameters can bistability occur even when condition
(\ref{bistability condition f=0}) breaks down. \ Note that unlike Eq.
(\ref{chi no cavity}), which only contains one critical point, Eq.
(\ref{depsilon}) can typically support two critical points. \ Thus, the
threshold for bistability in this cavity model can, in principle, occur at any
value of $n$, instead of always at $n=0.5$ as in the bare Feshbach model
discussed in the previous subsection. To proceed, we\ first change Eq.
(\ref{depsilon}) into a quartic equation for$\sqrt{y}$
\begin{equation}
\left(  \sqrt{y}\right)  ^{4}-\gamma\sqrt{y}+1=0,\label{y}%
\end{equation}
where $y=Cn-\delta$ and $\gamma=2\sqrt{B/\left[  \chi-h\left(  n\right)
\right]  }$. \ This change of variable is motivated by the realization that
when condition (\ref{bistability condition f=0}) breaks down or equivalently
$\chi+2\sqrt{2}>0$, the quantity $\chi-h\left(  n\right)  $ is always positive
\ so that only when $Cn-\delta>0$ can Eq. (\ref{depsilon}) holds. \ To
estimate the threshold condition at a given $n$, we regard Eq. (\ref{y}) as a
transcendental equation for $\delta$, and require Eq. (\ref{y}) to support a
real root of multiplicity 2 (another two roots are a complex conjugate pair).
\ This requirement allows us to conclude that bistability develops when
\begin{equation}
B\geq B_{th}\left(  n\right)  \equiv\frac{4}{3\sqrt{3}}\left[  \chi-h\left(
n\right)  \right]  ,\label{bistability condition with cavity}%
\end{equation}
and the bistability threshold at a given $n$ (and $C$) is reached when
$B=B_{th}\left(  n\right)  $ and $\delta=\delta_{th}\left(  n\right)  \equiv
Cn-1/\sqrt{3}$.%
\begin{figure}
[ptb]
\begin{center}
\includegraphics[
height=1.5093in,
width=3.6123in
]%
{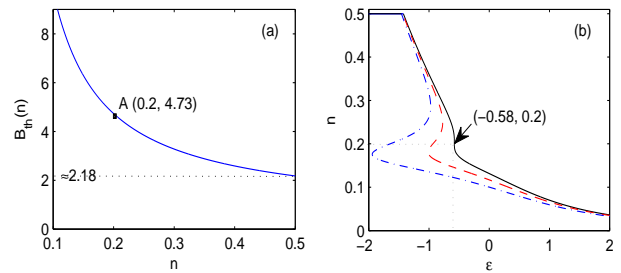}%
\caption{(Color online) (a) $B_{th}\left(  n\right)  $ as a function of $n$.
\ (b) Molecular population $n$ as a function of $\epsilon$ for $B=B_{th}%
\left(  0.2\right)  =$ $4.73$ (solid line), $B=2B_{th}\left(  0.2\right)  $
(dashed line), and $B=4B_{th}\left(  0.2\right)  $ (dot-dashed line) for a
cavity system with $C=20$, $\chi=0,$ and $\delta=\delta_{th}\left(
0.2\right)  =3.42$.}%
\label{Fig:bistabilityCavity}%
\end{center}
\end{figure}

$B_{th}\left(  n\right)  $ as a function of $n$ is displayed in Fig.
\ref{Fig:bistabilityCavity}(a). \ This simple threshold relation seems to hold
quite well as long as $C$ is sufficiently large. \ Consider, for example, a
cavity system with $C=20$. \ Figure \ref{Fig:bistabilityCavity}(b) shows how
$n$ changes with $\epsilon$ under different values of $B$. \ The solid line,
produced with $B$ $=B_{th}\left(  0.2\right)  =4.73$ [point A in Fig
\ref{Fig:bistabilityCavity}(a)] and $\delta=\delta_{th}\left(  0.2\right)
=3.42$, shows that the bistability threshold indeed occurs at the
theoretically predicted location with $n=0.2$ [and $\epsilon=-0.58$ determined
from Eq. (\ref{epsilon steady state})]. \ The dashed and dot-dashed lines in
Fig. \ref{Fig:bistabilityCavity}(b), produced with $B=2B_{th}\left(
0.2\right)  $ and $B=4B_{th}\left(  0.2\right)  $, respectively, clearly shows
that the usual bistable behavior emerges when $B$ is increased beyond its
threshold value $B_{th}\left(  0.2\right)  $. \ Recall that without a cavity,
a bistable transition can only take place between a mixed and a pure molecular
state. \ In the present situation with an optical cavity, we see that a new
feature appears - a bistable transition can also take place between two
different mixed states (the dashed line).

To see what condition (\ref{bistability condition with cavity}) means to the
cavity field, we consider a small-sized micro-ring cavity of total length
$\mathcal{L}=200$ $\mu m$ and finesse $\mathcal{F}=3.14\times10^{5}$. \ The
cavity is driven by an external laser tuned $\Delta_{a}=2\pi\times100$ GHz
away from the $D_{2}$ line, characterized with a wavelength $780$ nm and a
linewidth $2\pi\times3$ MHz, of sodium atoms confined to an effective spatial
region of $L=30\mu m$ and $A=\left(  10\mu m\right)  ^{2}$. \ In such a
cavity-condensate system, we have $\eta=2\pi\times0.0052$ MHz and $\kappa
=2\pi\times23.87$ MHz. \ Then, by using the same Feshbach resonance and the
atom number density in the previous subsection, we estimate that the minimum
number of photons that must be present inside the cavity to produce
bistability is $8.6\times10^{3}$. \ This figure is three orders of magnitude
smaller than the threshold photon number in a typical laser \cite{kimble94},
and can be further reduced with an appropriate choice of the system
parameters. \ 

Matter wave bistability at small photon numbers can be understood as follows.
\ In contrast to the phase shift due to the s-wave scattering, which is
linearly proportional to the molecular density, the phase shift, arising from
the feedback between optical and matter fields, is nonlinearly proportional to
the molecular density by way of Eq. (\ref{f scaled}) in a resonant fashion. On
one hand, the sensitivity (the change of this phase shift versus the change of
the molecular density) is bose-enhanced by the collective nature of the
condensate system. \ On the other hand, the peak of this phase shift for a
given cavity photon number can be significantly amplified in a microcavity
environment where photons are confined into a tiny volume. \ As a result,
molecular bistability is possible under a weak cavity field.

\section{Conclusion}

In this paper, we have studied the matter wave bistability in an intraspecies
Feshbach resonance model with and without the assistance of an optical cavity.
\ In particular, we have arrived at a set of conditions that allow the
bistability thresholds under the two different settings to be estimated
analytically. \ In the absence of a cavity, bistability is possible only when
the effective Kerr nonlinearity stemming from s-wave scatterings [Eq.
(\ref{bistability condition f=0})] is sufficiently negative. \ In the presence
of a cavity, even when condition (\ref{bistability condition f=0}) breaks
down, bistability can still occur, provided that $B$ in Eq. (\ref{B}), a key
parameter describing the cavity-mediated two-body interaction, is sufficiently
large [Eq. \ref{bistability condition with cavity}]. \ An important difference
between systems with and without a cavity is that the former supports one
stable mixed state while the latter can support two different stable mixed
states. (Both systems contain a pure molecular state.) Thus, a bistable
transition in the latter system can take place not only between a mixed state
and a pure molecular state as in the former model, but also between two
different mixed states.

\section{Acknowledgement}

This paper is dedicated to the memory of H. Y. L's former mentor Dr. Lorenzo
Narducci whose example gave new meaning to the words: \textquotedblleft
dedication\textquotedblright, \textquotedblleft hard work\textquotedblright,
\textquotedblleft impartiality", etc. \ This work is supported by US National
Science Foundation and US Army Research Office.%

\end{document}